\documentclass{aastex6}

\begin{document}


\title{Evidence of AGB pollution in Galactic globular clusters from the Mg-Al 
anticorrelations observed by the APOGEE survey}



\author{P. Ventura\altaffilmark{1}}

\author{D. A. Garc\'{\i}a--Hern\'andez\altaffilmark{2,3}}

\author{F. Dell'Agli\altaffilmark{1}}

\author{F. D'Antona\altaffilmark{1}}

\author{Sz. M{\'e}sz{\'a}ros\altaffilmark{4}}

\author{S. Lucatello\altaffilmark{5}}

\author{M. Di Criscienzo\altaffilmark{1}}

\author{M. Shetrone\altaffilmark{6}}

\author{M. Tailo\altaffilmark{1}}

\author{Baitian Tang\altaffilmark{7}}

\and

\author{O. Zamora\altaffilmark{2,3}}

\altaffiltext{1}{INAF -- Osservatorio Astronomico di Roma, via Frascati 33, I-00077 Monteporzio, Italy}
\altaffiltext{2}{Instituto de Astrof\'{\i}sica de Canarias (IAC), E-38205 La Laguna, Tenerife, Spain}
\altaffiltext{3}{Departamento de Astrof\'{\i}sica, Universidad de La Laguna (ULL), E-38206 La Laguna, Spain}
\altaffiltext{4}{ELTE Gothard Astrophysical Observatory, H-9704 Szombat- hely, Szent Imre Herceg st. 112, Hungary}
\altaffiltext{5}{INAF -- Osservatorio Astronomico di Padova, vicolo dell'Osservatorio 5, I-35122 Padova, Italy}
\altaffiltext{6}{University of Texas at Austin, McDonald Observatory, USA}
\altaffiltext{7}{Departamento de Astronom\'{\i}a, Casilla, 160-C, Universidad de Concepci{\'o}n, Concepci{\'o}n, Chile}

\begin{abstract}
We study the formation of multiple populations in globular clusters (GC), under the 
hypothesis that stars in the second generation formed from the winds of intermediate-mass
stars, ejected during the asymptotic giant branch (AGB) phase, possibly diluted with
pristine gas, sharing the same chemical composition of first-generation stars. To this aim,
we use the recent APOGEE data, which provide the surface chemistry of a large sample of
giant stars, belonging to clusters that span a wide metallicity range. The APOGEE data set
is particularly suitable to discriminate among the various pollution scenarios proposed so
far, as it provides the surface abundances of Mg and Al, the two elements involved in a
nuclear channel extremely sensitive to the temperature, hence to the metallicity of the
polluters. The present analysis shows a remarkable agreement between the observations and
the theoretical yields from massive AGB stars. In particular, the observed extension of the
depletion of Mg and O and the increase in Al is well reproduced by the models and the trend
with the metallicity is also fully accounted for. This study further supports the idea that
AGB stars were the key players in the pollution of the intra-cluster medium, from which
additional generations of stars formed in GC.
\end{abstract}

\keywords{Globular clusters: general --- stars: AGB and post-AGB --- stars: abundances}



\section{Introduction} \label{sec:intro}
The spectroscopic and photometric astronomical data accumulated in the last decades have
discredited the classic paradigma that GC stars represent a  simple stellar population.
Results from high-resolution spectroscopy outlined star-to-star differences in the
abundances of the light elements, suggesting the signature of proton-capture
nucleosynthesis \citep{gratton12}. The discovery of multiple main sequences (MS) in the
color-magnitude  diagrams of some GC \citep{piotto07} indicated the existence of a group of
stars  enriched in He, thus further confirming the hypothesis that part of the stars in  GC
formed from proton-capture contaminated matter. 

Several self-enrichment mechanisms have been proposed so far to explain the formation of
multiple populations in GC: the AGB scenario \citep{ventura01,  dercole08}, fast rotating
massive stars \citep{decressin07}, interacting massive binaries \citep{demink09}, core
H-burning in supermassive MS stars \citep{denissenkov14} and the accretion model
\citep{bastian13}. The difficulties encountered by the different scenarios in accounting
for the various observational evidences have been recently discussed in \citet{renzini15}.

Until now, the theoretical attempts to reproduce the abundance patterns observed in GC 
stars have been mainly focused on the O-Na anticorrelation (as obtained from 
high-resolution optical spectroscopy), which has been detected in essentially all Galactic 
GC where the compositions for more than a handful of stars have been measured \citep[see
e.g.][]{carretta09}. High-resolution near-IR massive spectroscopic surveys  like the Apache
Point Observatory Galactic Evolution Experiment \citep[APOGEE,][]{majewski16} are going to
strongly augment these previous optical studies. This is because APOGEE provides: i) the
chemical abundances of key elements  like Fe, Mg, Al, etc. from neutral H-band lines that
are expected to be less affected by nonlocal thermodynamic equilibrium effects than those
in the optical range  \citep[see e.g.][]{garcia15}; and ii) a higher number of GC stars 
(compared to literature) analysed in a homogeneous way, and covering GC of  various mass
and metallicity \citep{meszaros15}. Indeed, the Mg-Al distributions are much clearer in
the APOGEE data \citep{meszaros15} than in previous optical studies
\citep[e.g.][]{carretta13}.

The analysis of the Mg-Al anticorrelation offers a better opportunity to deduce the nature 
of the polluters in comparison to the traditional O-Na trend, for the following reasons: a)
the activation of the Mg-Al nucleosynthesis requires higher temperatures than the
CNO and NeNa cycles, thus allowing to further restrict the range of  temperatures at which
the contaminated matter was exposed; b) the Mg and Al abundances of red giant branch (RGB)
stars definitively reflects the initial chemical composition, as no change in the surface
abundance of these elements  occurs during the RGB phase; unlike O and Na, whose surface
content can be affected by deep mixing \citep{denisenkov90}.

Here we interpret the recent APOGEE data of GC stars \citep{meszaros15}, which clearly show
that the matter from which second-generation (SG) stars formed was depleted  in Mg. This
finding rules out fast rotating massive stars and massive binaries as possible polluters,
as well as the accretion scenario hypothesis\footnote{We do not discuss the supermassive MS
stars hypothesis, because no self-consistent model based on hydrodynamic simulations has
been presented so far and it is not yet clear if and how the nucleosynthesis associated 
with this mechanism is sensitive to metallicity.}. We concentrate on the AGB  scenario, in
which SG stars in GC formed from the winds of massive AGB stars (AGBs), mixed with  
pristine gas in the
cluster. We base our analysis on five GC, selected to span a wide range  of metallicities
and with a number of stars observed sufficiently large to allow a  statistical
interpretation of the results obtained.

To date, this is the first investigation where the AGB scenario is tested against an 
extensive data set of Mg and Al abundances. This study is the most severe 
and reliable test of the role that AGB stars may have played in the self-enrichment 
of GC, as it allows to fix with an unprecedent accuracy the extent of the
nucleosynthesis to which the contaminating matter was exposed and, more importantly,
how this changes with metallicity.

\section{Pollution from massive AGB stars}

Pollution from stars of mass above $\sim 3~M_{\odot}$ is determined by hot bottom burning
(HBB), the nuclear activity taking place at the base of the convective envelope (CE) during
each interpulse \citep{blo91}. The temperature at the bottom of the CE,
$T_{bce}$, is the key quantity in determining the degree of the nucleosynthesis
experienced and the chemical composition of the gas ejected into the interstellar medium.
The activation of CN cycling takes place when $T_{bce} \sim 40$  MK, whereas O burning and
Na production (via $^{22}$Ne proton capture) require $T_{bce} \sim 80$ MK; at $T_{bce} \sim
100$ MK Mg is efficiently destroyed by proton fusion, starting a series of reactions
leading to the synthesis of Al. At even higher $T_{bce}$ part of the synthesised Al
produces Si, via proton capture.

\begin{figure}
\figurenum{1}
\plottwo{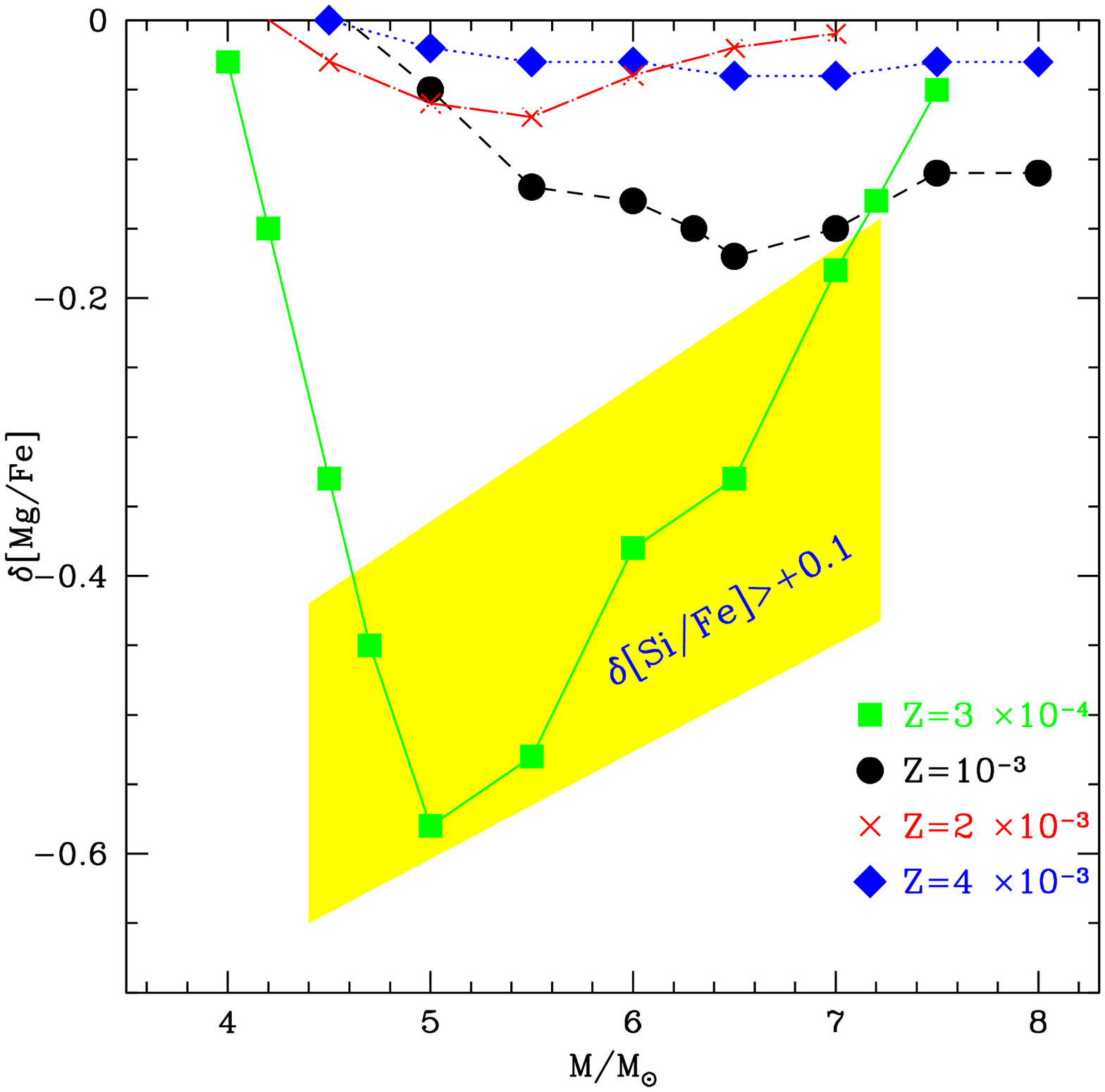}{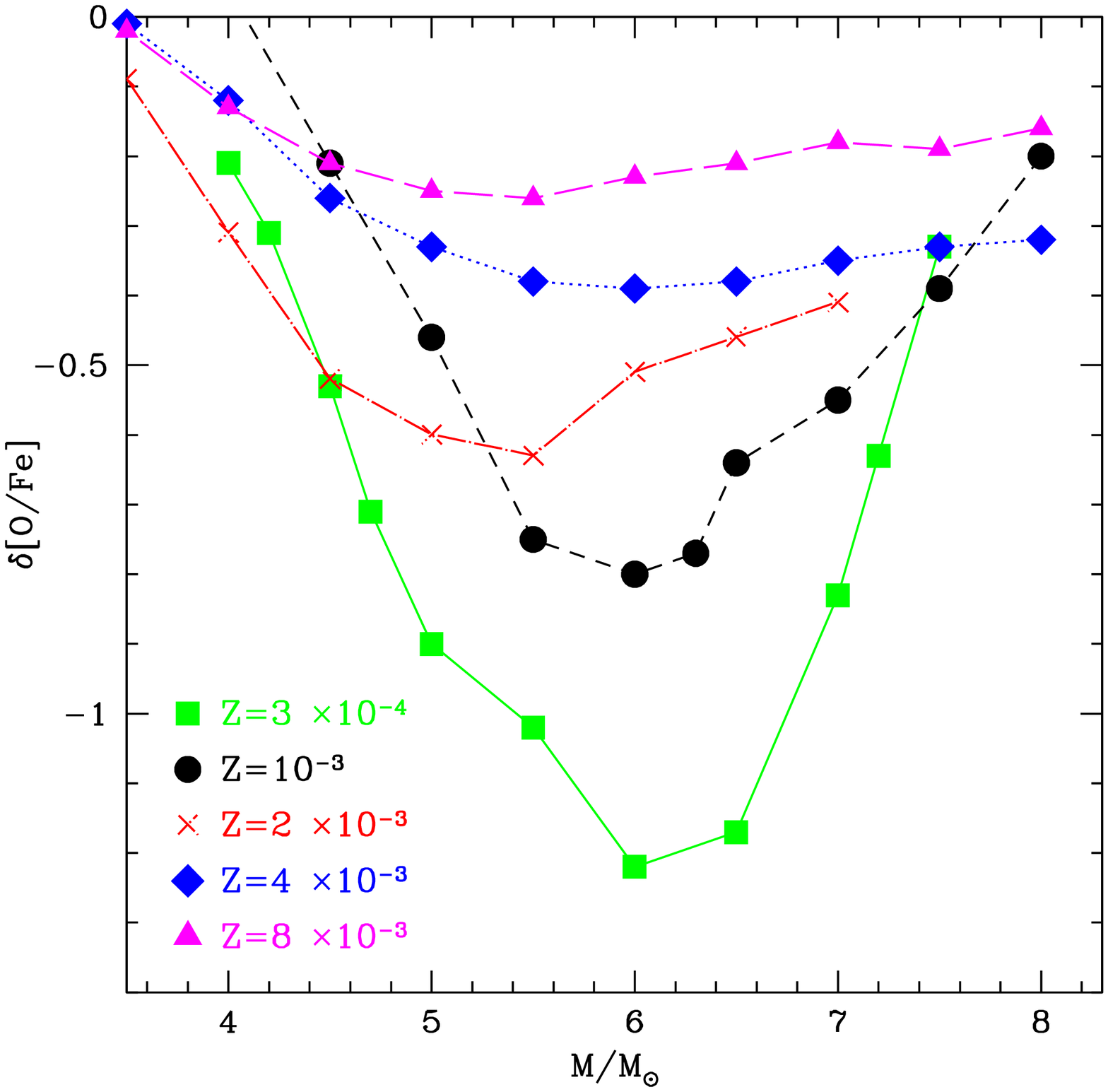}
\caption{The average depletion of Mg (left) and O (right) in models
of different mass and metallicity. The meaning of the symbols is as follows:
green squares, black points and magenta triangles indicate, respectively, 
$Z=3\times 10^{-4}$, $Z=10^{-3}$ and $Z=8\times 10^{-3}$ models by
\citet{ventura13}; blue diamonds and red crosses correspond, respectively, to
$Z=4\times 10^{-3}$ models by \citet{ventura14} and to unpublished, 
$Z=2\times 10^{-3}$ models. The yellow, shaded area in the left panel includes
the models that produce a Si enrichment above 0.1 dex. The $Z=8\times 10^{-3}$ 
models are not shown in the left panel, as no significant Mg depletion occurs.
\label{figdelta}}
\end{figure}

The nucleosynthesis activated at the base of the CE is extremely sensitive to the mass and
the metallicity: a) stars of higher mass evolve on more massive cores, thus reaching
higher temperatures and experiencing a more advanced nucleosynthesis; b) lower metallicity
stars evolve at larger $T_{bce}$, thus they experience a more efficient proton-capture
burning. 

The depletion of Mg and O in the gas lost by AGB stars in models of various mass and 
metallicity is shown in Fig.~\ref{figdelta}. These results refer to  
$Z=3\times 10^{-4}$, $Z=10^{-3}$ and $Z=8\times 10^{-3}$ models by \citet{ventura13}, to 
$Z=4\times 10^{-3}$ models by \citet{ventura14} and to unpublished models of metallicity 
$Z=2\times 10^{-3}$. In all cases mass loss was modelled according to \citet{blo95},
where $\dot M \sim L^{4.2}/T_{eff}^2$, with no explicit dependency on the surface 
chemistry.

Basically, we find the following: a) CN cycling is activated in the whole range  of mass
and metallicity explored here; b) O destruction occurs in stars of mass above $\sim
4~M_{\odot}$, with efficiency decreasing with increasing $Z$; c) Na is synthesized via
$^{22}$Ne-burning in the same range of masses at which O destruction takes place; d) Mg
depletion occurs only in low-metallicity massive AGB  stars; e) Si is produced only in
$Z<5\times 10^{-4}$ massive AGBs.

Fig.~\ref{figdelta} shows that Mg depletion, compared to O-burning, is achieved only in
low-metallicity AGBs. Furthermore, the synthesis of Al is extremely sensitive to
metallicity, at odds with Na production, which, for the reasons given above, is effective
in all the AGB models of mass above $\sim 4~M_{\odot}$, independently of $Z$. It is for these
reasons that the investigation of the Mg-Al anticorrelation is a much more efficient tool,
compared to the classic O-Na trend, to explore and test all the predictions of the AGB
scenario for the formation of multiple populations in GC.

\section{Understanding the chemical patterns of GC stars}

\floattable
\begin{deluxetable}{ccccccc}
\tablenum{1}
\tablecaption{The chemical properties of Globular Clusters \label{tabGC}}
\tablewidth{0pt}
\tablehead{
\colhead{GC ID} & \colhead{GC Name} & \colhead{[Fe/H]} & \colhead{$\Delta[Si/Fe]$} & 
\colhead{$\Delta[Al/Fe]$} & \colhead{$\Delta[Mg/Fe]$} & \colhead{$\Delta[O/Fe]$}
}
\startdata
NGC 6341 &  M92  & $-2.23$ & 0.06 & 0.93 & 0.28 & 0.27 \\
NGC 6205 &  M13  & $-1.50$ & 0.02 & 0.95 & 0.12 & 0.36 \\
NGC 5272 &  M3   & $-1.40$ & 0.01 & 0.87 & 0.08 & 0.23 \\
NGC 5904 &  M5   & $-1.24$ &  -   & 0.70 & 0.03 & 0.25 \\
NGC 6171 &  M107 & $-1.01$ &  -   & 0.02 &  -   & 0.09 \\
\enddata
\tablecomments{The entries in columns 4-7 give the spread
between the average mass fractions of Si, Al, Mg and O in stars belonging to the FG
and SG of the cluster, as identified by \citet{meszaros15}}
\end{deluxetable}

In the interpretation of the APOGEE data we focus on M92, M3, M13, M5 and M107. 
This choice is motivated by the wide spread in metallicity spanned by the stars in 
these clusters and, for the most metal-poor GC, by the large number of giants observed. 
This allows, at least on qualitative grounds, an approach aimed at understanding how star 
formation occurred in these stellar systems. Although the APOGEE data include C and N here 
we focus on Mg, Si, Al and O, because: a) CN cycling is much less sensitive to metallicity 
compared to the nuclear channels involving the heavier species; b) C and N are subject to 
possible alterations during the RGB evolution, which prevents a straightforward interpretation of 
the measured abundances. The data of the clusters used in the present analysis are reported 
in Table \ref{tabGC}. We discuss the individual clusters separately, starting with the 
most metal-poor GC M92 and ending with the most metal-rich, M107.

\begin{figure}
\figurenum{2}
\plottwo{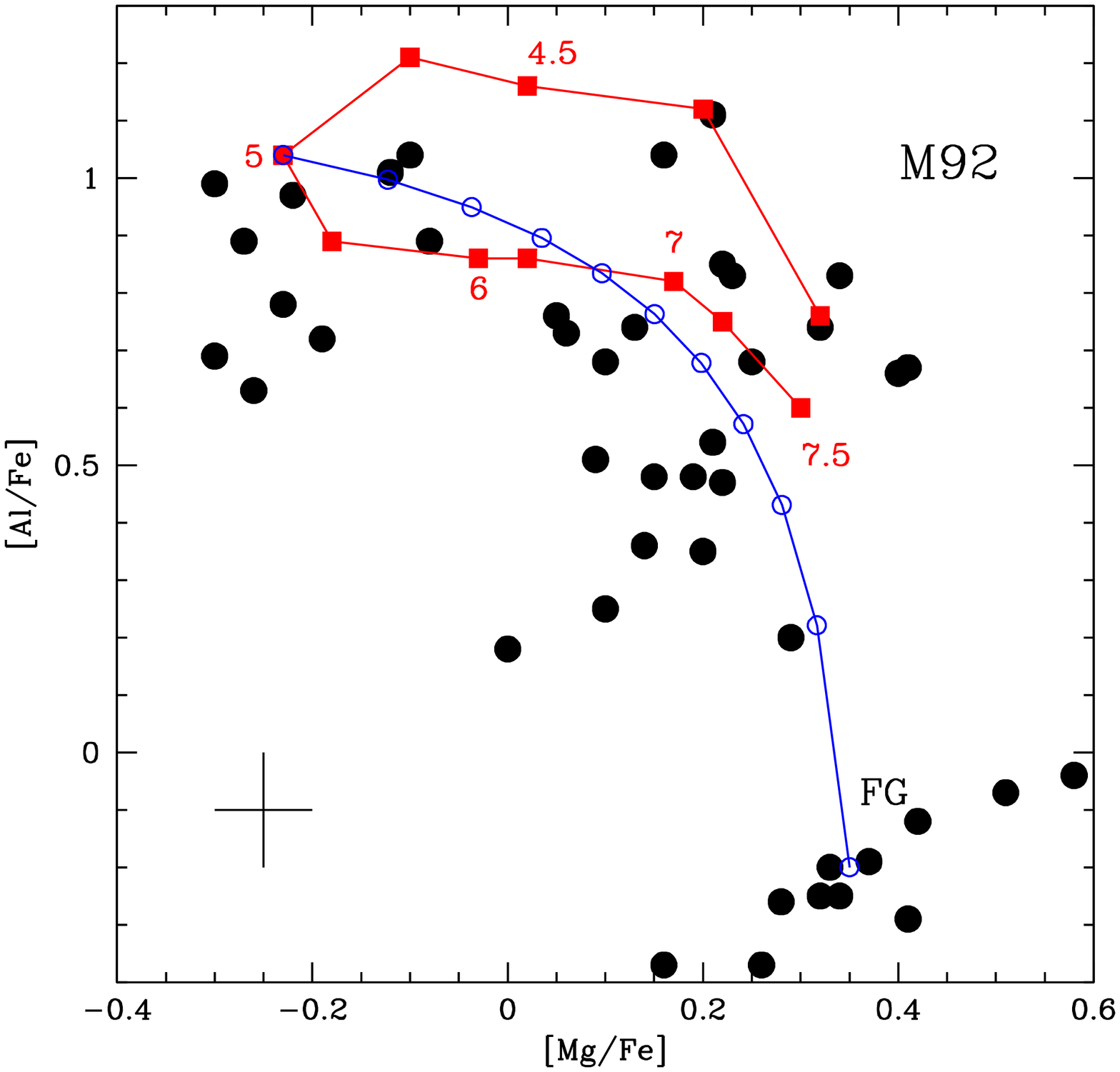}{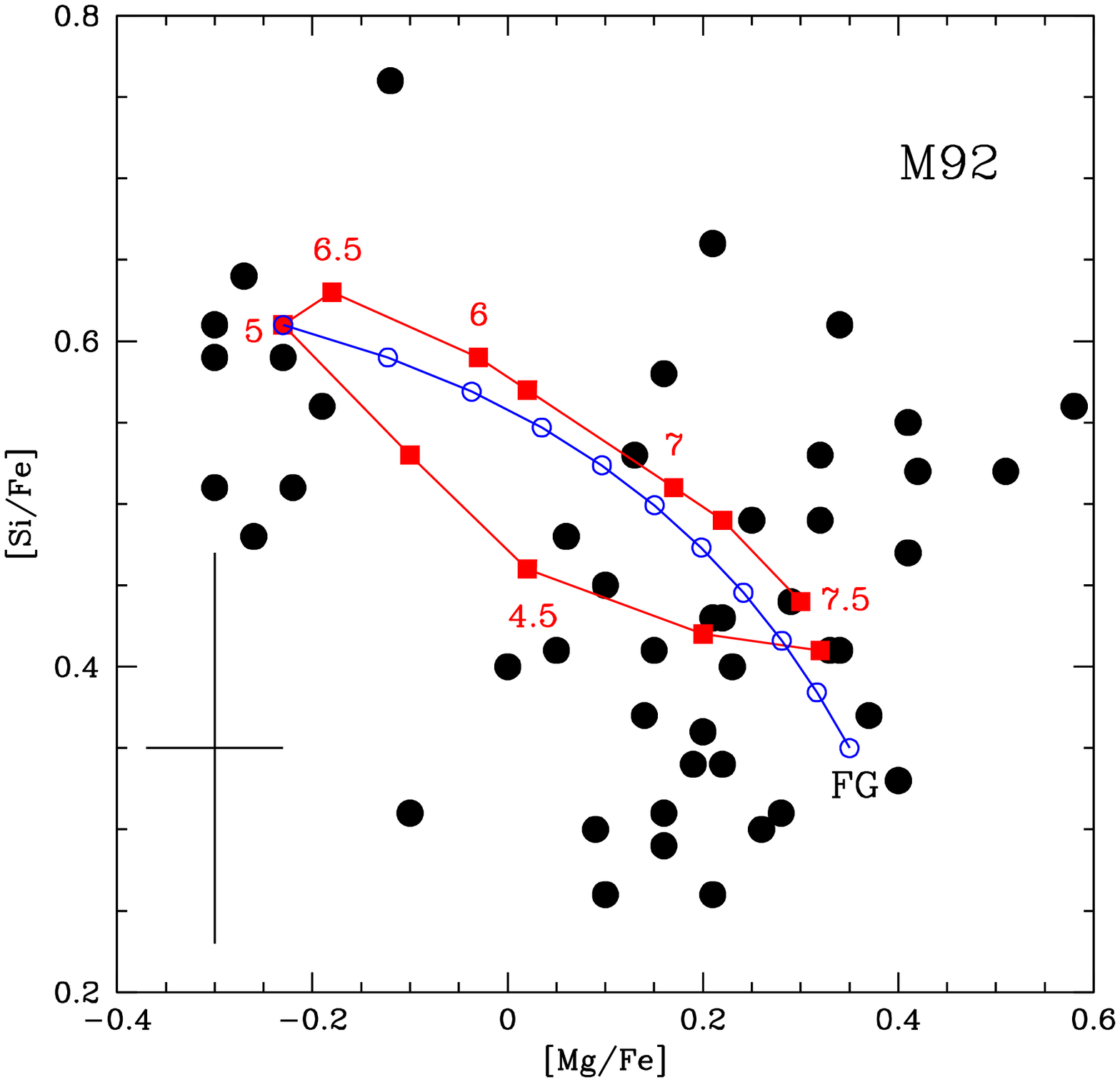}
\caption{M92 stars from \citet{meszaros15} are shown as full dots in the Mg-Al (left)
and Mg-Si (right) planes. The black crosses indicate typical error bars.
The yields of AGB models of metallicity $Z=3\times 10^{-4}$
are indicated with red squares; the numbers close to the squares indicate the initial
masses. Blue, open circles indicate the chemistry expected from
mixing of the AGB ejecta with pristine gas, with percentages of the latter ranging
from 0 to $100 \%$, in steps of $10 \%$. The points corresponding to $100 \%$ dilution 
are close to the FG label.
\label{figm92}}
\end{figure}

\begin{figure}
\figurenum{3}
\gridline{\fig{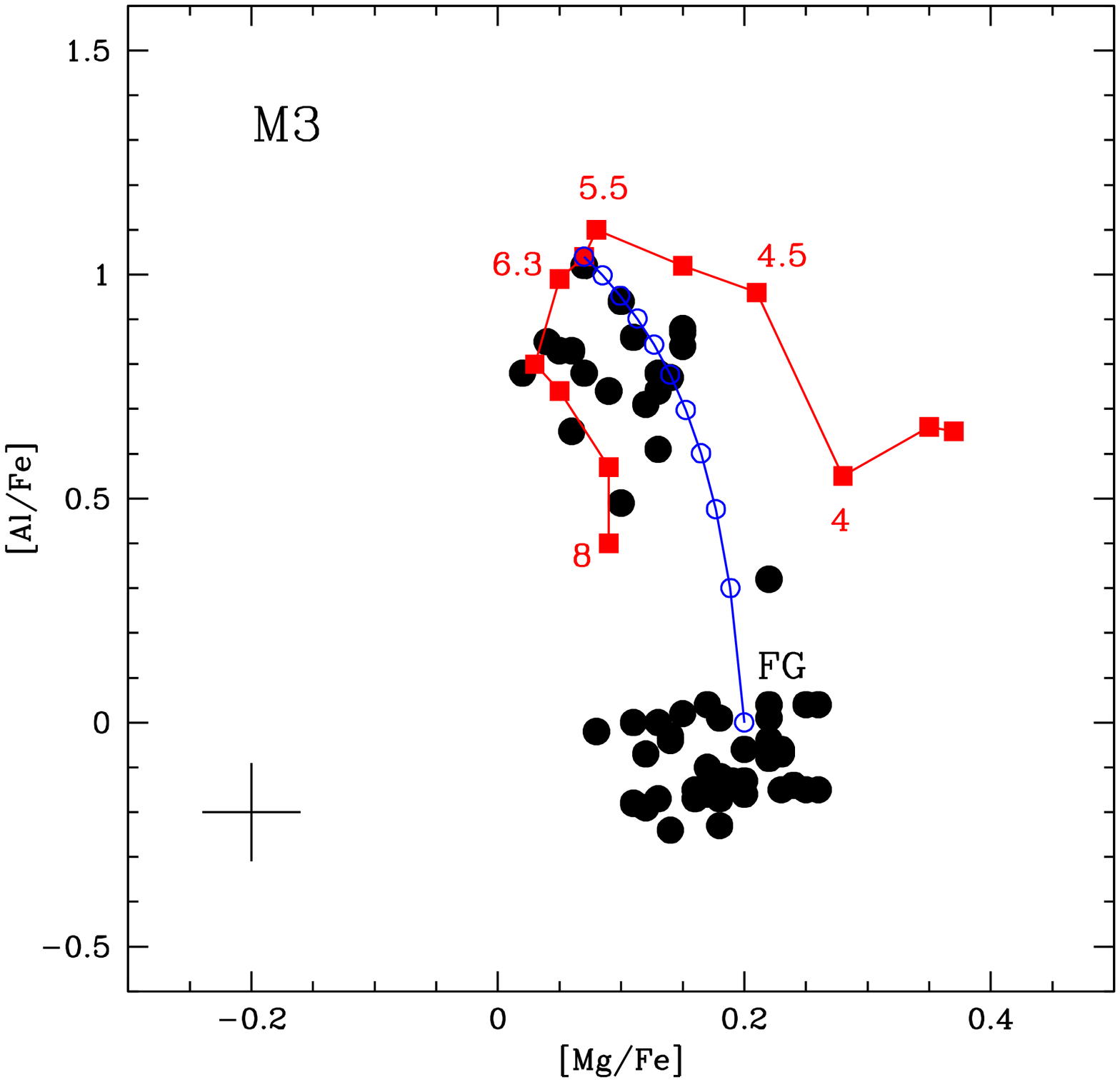}{0.42\textwidth}{(a)}
          \fig{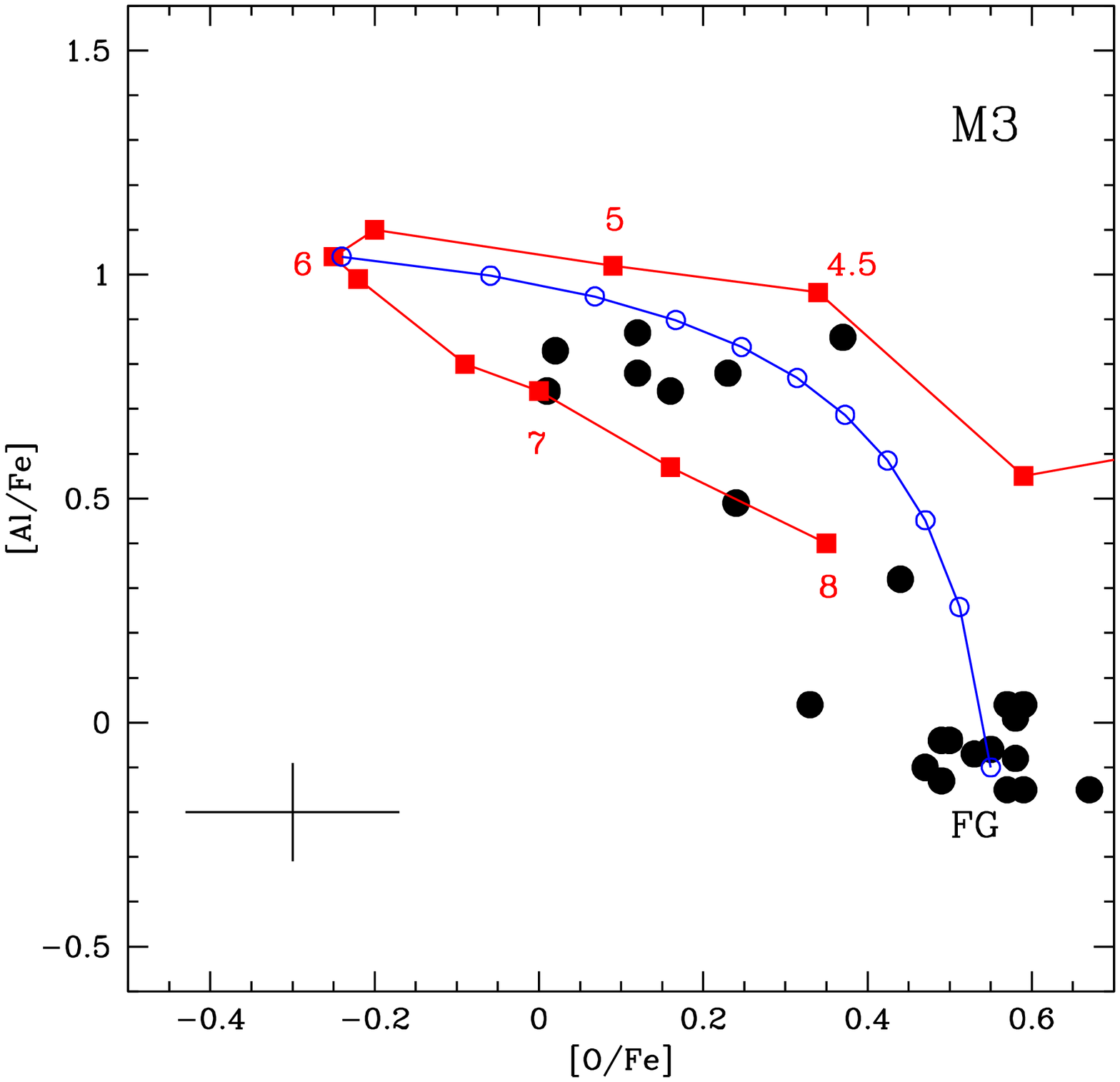}{0.42\textwidth}{(b)}
          }
\gridline{\fig{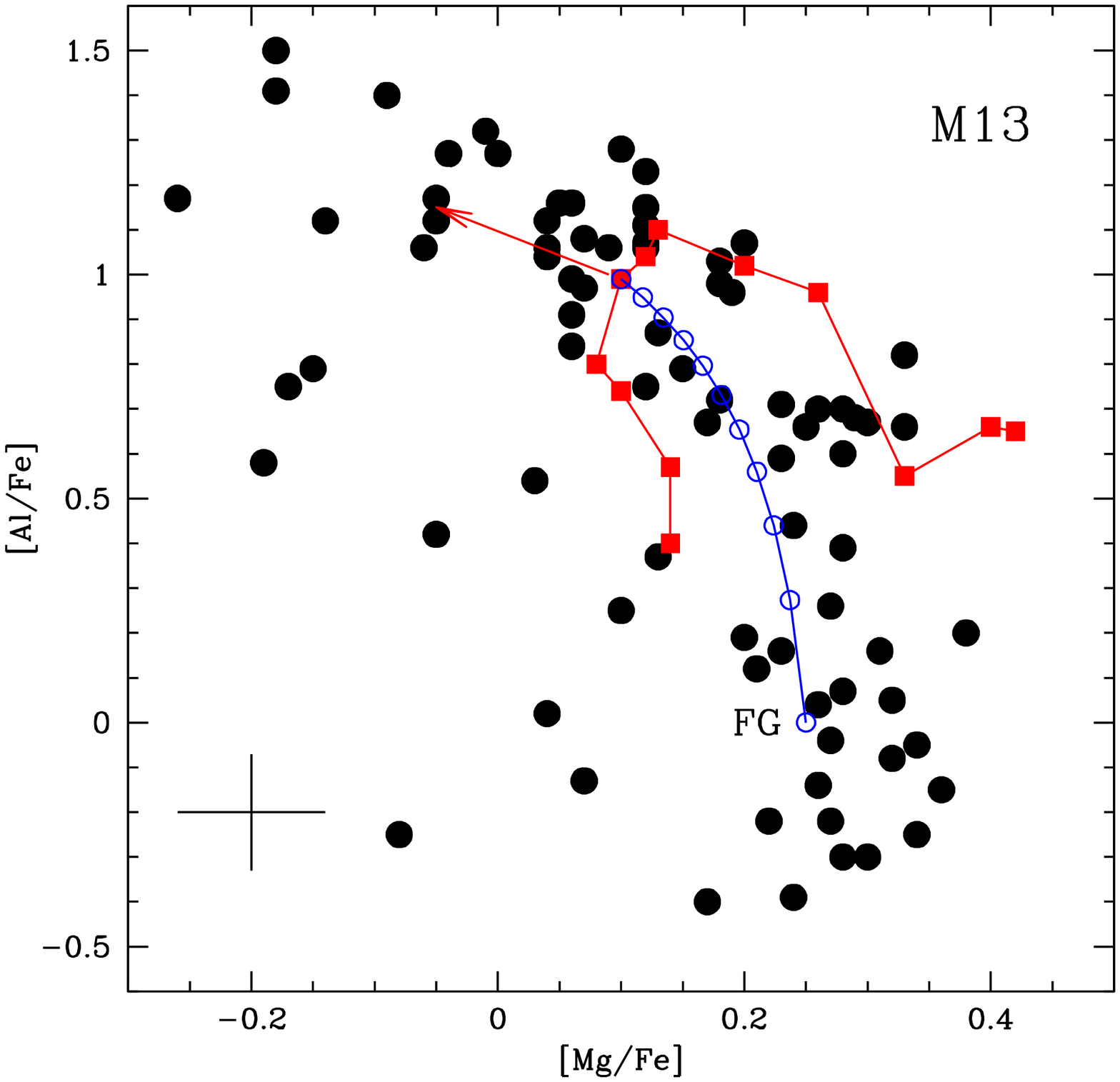}{0.42\textwidth}{(c)}
          \fig{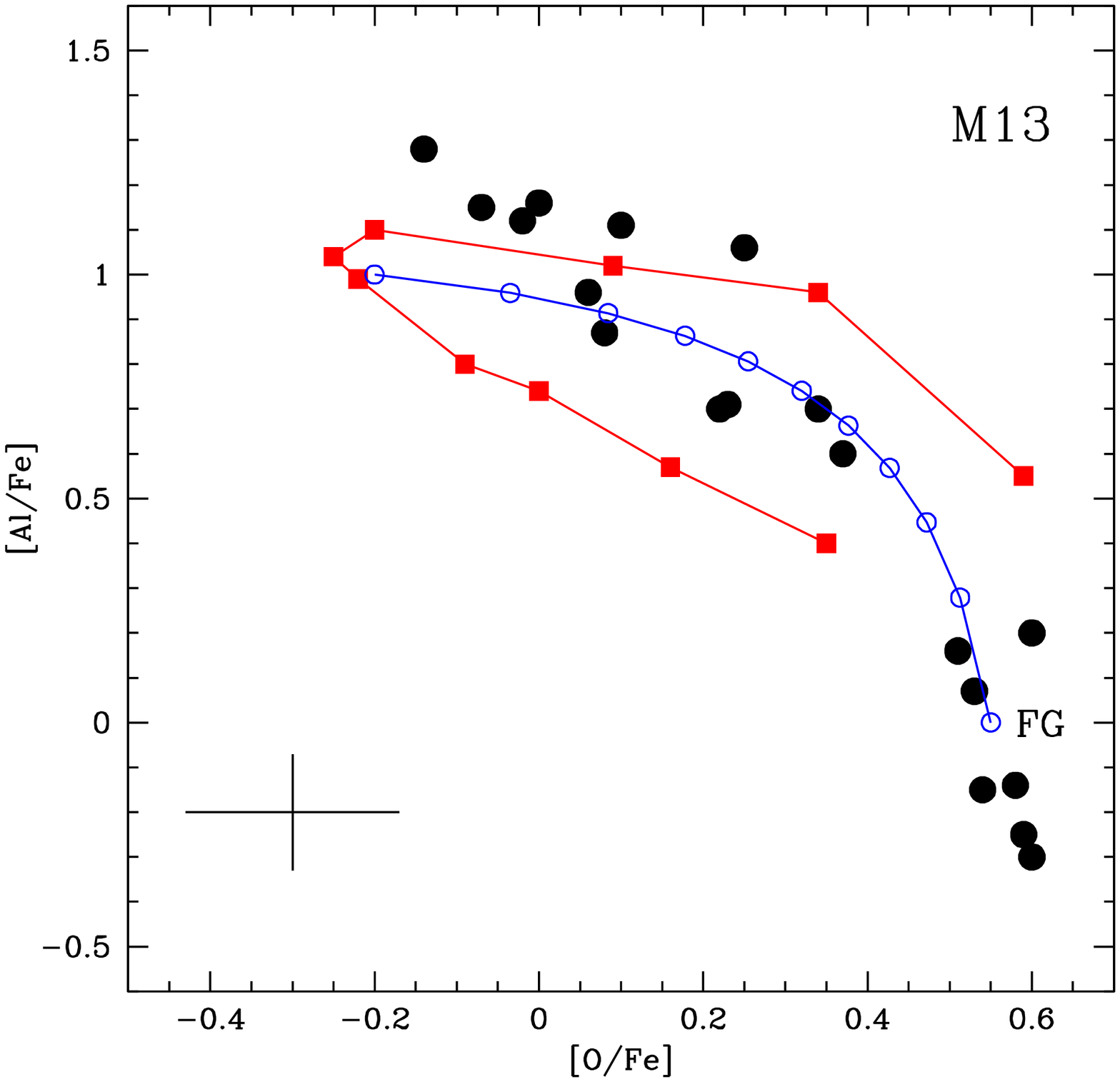}{0.42\textwidth}{(d)}
          }
\caption{The Mg-Al (left) and O-Al (right) trends defined by stars in M3 (top) and
M13 (bottom). The observations are compared with AGB models of metallicity
$Z=10^{-3}$. The meaning of the symbols is the same as in Fig.~\ref{figm92}. 
\label{figm3}}
\end{figure}

\subsection{M92}

M92 is the lowest metallicity GC in our sample. Fig.~\ref{figm92} shows the observed
Mg-Al and Mg-Si trends; the observations are compared with the $Z=3\times 10^{-4}$ models
shown in Fig.~\ref{figdelta}. On the same planes we also show a dilution curve, obtained by
mixing different fractions of the AGB ejecta with pristine gas. To have an idea of the
largest degree of contamination expected, here and in the following the dilution patterns
are based on the chemistry of the AGB models showing the largest modification with respect
to the initial chemistry;  \citet{vd11} showed that this is found for the stars whose mass
is just below the threshold required to activate C-burning and develop a core
composed of O and Ne. 

The chemical composition of AGBs nicely reproduces the observations, particularly  the
spread in Mg (0.6 dex), Al (1 dex) and Si (0.2 dex). These results confirm that the 
contaminated matter was exposed to a very advanced nucleosynthesis, above  $\sim 100$ MK,
in agreement with the predictions of low-metallicity AGB models.

According to our interpretation, the stars with the most contaminated chemical 
composition, in the left-upper regions of the panels of Fig.~\ref{figm92}, formed from
the ashes of the evolution of massive AGBs, with little dilution with pristine gas.

The sample of M92 giants shown in Fig.~\ref{figm92} is completed by first generation (FG) 
stars, sharing the same chemical composition of the pristine gas, and an additional group 
of giants, whose chemistry invokes dilution of the AGB winds with $30-70 \%$ of pristine 
gas. This suggests that the formation of SG stars in M92 took place as a rather continuous
process, though possible gaps in the formation of the SG cannot be ruled out based on the 
present data.

\subsection{M3 and M13}

The Mg-Al and O-Al trends traced by M3 and M13 stars are shown in Fig.~\ref{figm3}, where
they are compared with the ejecta of $Z=10^{-3}$ AGBs. 

The models reproduce the position on the Mg-Al and O-Al planes of the M3 stars with the
most contaminated chemical composition. Compared to M92, a narrower Mg distribution  is
found here, consistently with the higher metallicity of M3 and with the results shown in 
Fig.~\ref{figdelta}. The largest O-depletion observed is $\sim 0.2-0.3$ dex lower than the
AGB yields, suggesting that dilution with pristine gas occurred in the formation of the
SG; conversely, the observed Mg depletion of the same group of objects is well reproduced
by the models. These results might suggest that the predicted Mg depletion is slightly
underestimated at these metallicities.

The right-lower panel of Fig.~\ref{figm3} shows that the observed O-Al trend of M13 is
nicely reproduced: the M13 stars with most extreme chemical composition show O-depletion
($\delta[O/Fe]=-0.6$) and Al-enhancement ($\delta[Al/Fe] \sim 1$) in agreement with the
theoretical predictions.

In the Mg-Al plane a few stars show a Mg-depletion ($\sim$0.2 dex) in excess of the 
theoretical expectations, once more suggesting the need for a slightly higher Mg-depletion.
This problem was addressed by \citet{dantona16} and \citet{ventura11},
who discussed the possibility that a factor of $\sim$2 increase in the cross-section of the
$^{25}$Mg(p,$\gamma)^{26}$Al reaction was required to fill this gap. The expected results,
indicated with an arrow in the left-bottom panel of Fig.~\ref{figm3}, confirm that the
agreement with the data would be significantly improved. We leave this problem open.

In the relative comparison between the two clusters, the distributions of stars in the
various planes indicate that, on average, the SG of M13 is more contaminated by the AGBs
ejecta, whereas in the case of M3 a higher degree of dilution with pristine gas occurred.

\subsection{M5}

The Mg-Al trend defined by M5 stars is rather peculiar, as there is no appreciable Mg 
spread ($\delta[Mg/Fe]<0.03$), while the Al-spread is about a factor of 10 
\citep{meszaros15}. This can  be understood based on the behavior of the $Z=2\times 10^{-3}$ 
lines in Fig.~\ref{figdelta}. Indeed, the M5 metallicity is just above the threshold 
required to achieve a significant
Mg-depletion, but still sufficiently low that Al-synthesis occurs (this apparently
anomalous behavior is due to the large difference in the initial mass fractions of the two
elements, such that the consumption of a small percentage of Mg is sufficient to allow a
significant Al production).

\begin{figure}
\figurenum{4}
\plottwo{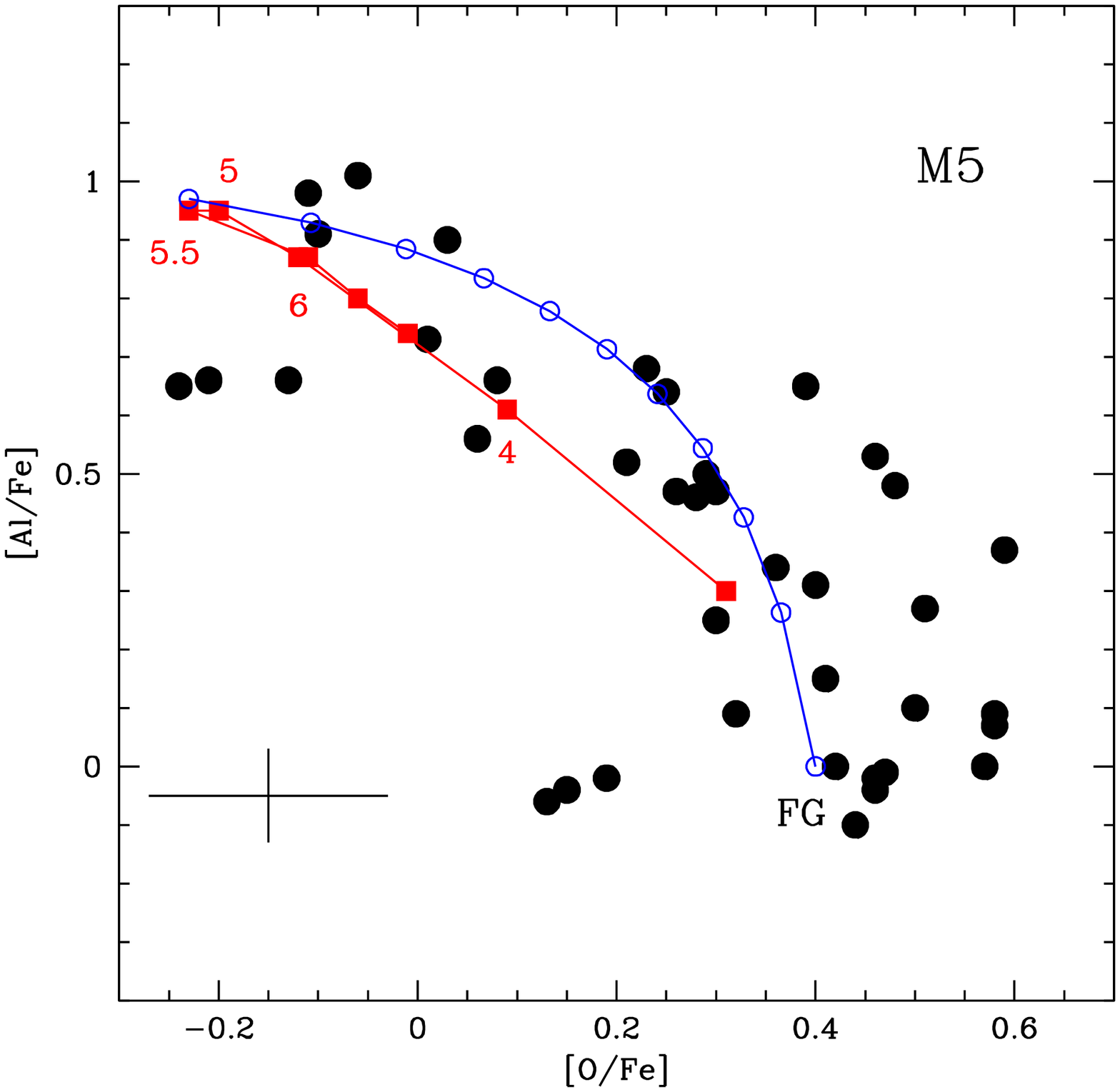}{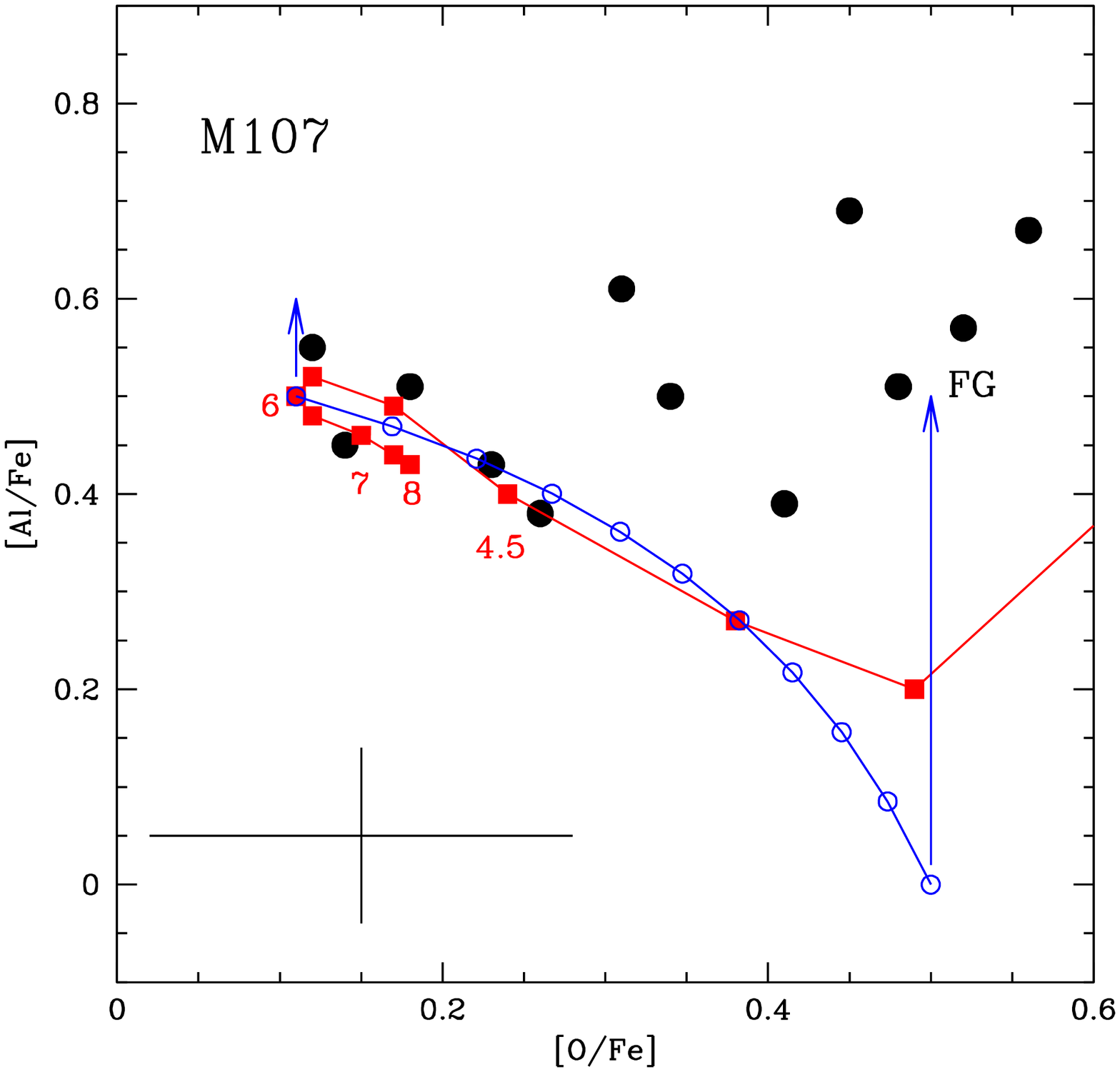}
\caption{The O-Al trends defined by stars in M5 (left) and M107 (right),
compared with AGB models of metallicity, respectively, $Z=2\times 10^{-3}$ and
$Z=4\times 10^{-3}$. The arrows on the right indicate the change in the dilution
curve when the higher Al of FG stars in M107 is considered. 
\label{figm5}}
\end{figure}

The results in the O-Al plane (left panel of Fig.~\ref{figm5}) shows that the chemistry of
the AGB ejecta nicely compares with the chemical composition of the SG stars: the largest
O-depletion ($\delta[O/Fe]=-0.5$) and Al-increase ($\delta[Al/Fe]=1$) are nicely reproduced
by the AGB models. The difference between the O spread defined by M5 
($\delta[O/Fe]=-0.5$) and M13 ($\delta[O/Fe]=-0.7$) stars is in agreement with the results
from AGB models of the metallicity of the two clusters, as can be deduced from the comparison
between the $Z=10^{-3}$ and $Z=2\times 10^{-3}$ lines in Fig.~\ref{figdelta}.

The observations apparently trace a continuous trend, suggesting that SG stars in M5 formed
from the gas ejected from massive AGBs diluted at various extents with pristine
matter.

\subsection{M107}

M107 is the most metal-rich cluster examined here. Although the number of stars with 
available O abundances \citep{meszaros15} is not as large as in the other clusters,  we
preferred to discuss it, because this allows a wider exploration of the  metallicity
effects. 

No Mg spread is detected, a consequence of the poor efficiency of Mg-burning in AGB stars
of metallicity $Z\geq 4\times 10^{-3}$. The O-Al trend of M107 is shown in the right panel
of Fig.~\ref{figm5}. 

The O abundances measured in M107 stars span a range of $\sim$0.4 dex, significantly
smaller than observed in the GC discussed so far.  While the small number of M107 stars 
with measured O abundances might partly account for this difference, it is however hard to 
believe that stars with O abundances below $[O/Fe]=-0.1$ can be detected. Such a small
spread is motivated by the relative high metallicity of M107, because O-burning becomes
less and less efficient as the metallicity increases. This can be seen in the right panel
of Fig.~\ref{figdelta}, where we note that the line corresponding to $Z = 4\times 10^{-3}$
never drops below $\delta [O/Fe]=-0.3$. 

The Al abundances are not of help in this case. Despite the models predicting some Al
production ($\delta [Al/Fe] \sim +0.5$), no significant spread is observed, because the FG
stars in this cluster, those on the right side of the O-Al plane, with $[O/Fe] \sim
0.4-0.5$, formed with a higher Al ($[Al/Fe]=+0.5$), compared to a purely solar scaled
mixture; in this case, for the same increase in the overall Al, no significant 
percentage increase of the surface Al is expected, in agreement with the observations.

\section{Conclusions}
We discuss the AGB scenario for the formation of multiple populations in GC, based on
recent results from the APOGEE survey. This investigation is focused mainly on the
interpretation of the Mg-Al anti-correlation, which compared to the largely used O-Na
trend, offers a much better tool to identify the potential polluters of the intra-cluster
medium. This is due to the much higher sensitivity of Mg-Al cycling to temperature and
metallicity. Furthermore, the interpretation of the Mg and Al content of
giant stars is straightforward, since both species are not expected to undergo any
alteration during the RGB evolution, even in the case of strong extra-mixing events.

To fully exploit the possibilities offered by the APOGEE observations, we examine five
clusters, spanning the metallicity range $-2.3 < [Fe/H] < -1.0$. This is the first time
that a self-enrichment scenario is tested against the O-Mg-Al data in a large sample of GC
stars spanning such a wide (a factor $\sim 20$) metallicity range.

The comparison between the data and the chemical composition of the ejecta of massive 
($M \geq 4~M_{\odot}$) AGB stars shows a remarkably good agreement. The observed
O-Mg-Al spreads are reproduced and more importantly, the AGB yields follow the
same trend with metallicity outlined by the observations, derived by comparing results of
clusters with different metal content. In all the GC examined, dilution of AGB gas
with pristine matter is required to match the observations. Part of SG stars
in M92 and M13 appear to have formed from pure AGB ejecta, with no dilution.

The present analysis provides a strong indication
that the nucleosynthesis experienced by AGB stars is characterised by the large variety of
temperatures, changing according to the mass and the metallicity, required to reproduce the
chemical patterns traced by GC stars.

\acknowledgments
PV acknowledges the Instituto de Astrof\'{\i}sica de Canarias for inviting him as a Severo Ochoa
visitor during 2016 August when most of this work was done. DAGH was funded by the
RYC$-$2013$-$14182 fellowship and DAGH/OZ acknowledge support provided by the MINECO grant
AYA$-$2014$-$58082-P. SM has been supported by the Premium Postdoctoral
Research Program of the Hungarian Academy of Sciences, and by the
Hungarian NKFI Grants K-119517 of the Hungarian National Research, Development
and Innovation Office. Funding for SDSS-III has been provided by the Alfred P. Sloan
Foundation, the Participating Institutions, the National Science Foundation, and the U.S.
Department of Energy Office of Science. The SDSS-III web site is http://www.sdss3.org/.
SDSS-III is managed by the Astrophysical Research Consortium for the Participating
Institutions of the SDSS-III Collaboration including the University of Arizona, the
Brazilian Participation Group,  Brookhaven  National  Laboratory,  University  of
Cambridge, Carnegie Mellon University, University of Florida, the French Participation
Group, the German Participation Group, Harvard University, the Instituto de Astrofísica de
Canarias, the Michigan State Notre Dame JINA Participation Group, Johns Hopkins University,
Lawrence Berkeley National Laboratory, Max Planck Institute for Astrophysics, New Mexico
State University, New York University, Ohio State University, Pennsylvania State
University, University of Portsmouth, Princeton  University,  the  Spanish  Participation 
Group, University of Tokyo, University of Utah, Vanderbilt University, University of
Virginia, University of Washington, and Yale University.

\vspace{5mm}
\facilities{SDSS-III(APOGEE)}

\software{ATON}


\listofchanges

\end{document}